\documentclass[12pt]{article}
\begin{document}

\title{Intrinsic Third Order Aberrations in Electrostatic and Magnetic
  Quadrupoles}
\author{R.~Baartman, TRIUMF, \\4004 Wesbrook Mall, Vancouver, B.C., Canada V6T 2A3}
\date{March, 1997}
\maketitle

\begin{abstract}  
Intrinsic aberrations are those which occur due to the finite length of the
desired field configuration. They are often loosely ascribed to the
fringing field. This is misleading as it implies that the effects can be
minimized by shaping the fields. In fact, there is an irreducible component
related to the broken symmetry. It is present even in the hard-edge limit,
and moreover, the other (soft-edge) effects can be simply ascribed to the
intrinsic aberration spread over a finite length.

We rederive the aberration formulas for quadrupoles using a Hamiltonian
formalism. This allows for an easy comparison of electrostatic and magnetic
quadrupoles. For different combinations of large and small emittances in
the two transverse planes, it is found that in some situations
electrostatic quadrupoles have lower aberrations, while in others, magnetic
quadrupoles are better.  As well, we discuss the ways in which existing
transport codes handle quadrupole fringe fields. Pitfalls are pointed out
and improvements proposed.
\end{abstract}

\section{Introduction}
A common prescription for quadrupole design is that the beam occupy no more
than a certain fraction, usually 1/2 to 3/4, of the aperture. Another
common prescription is that the longer the quadrupole is compared with its
bore diameter, the better. As well, it is common practice to carefully
round the ends of the poles.  We show that none of these common practices
can be validated by the quadrupole dynamics up to 3$^{\rm rd}$ order in
force.

The misconceptions are perpetuated by existing transport codes like {\tt
  GIOS}\cite{mw} and {\tt COSY}\cite{berz}, because they allow one to make
calculations with no fringe fields. Then when fringe fields are included,
aberrations increase. In fact, the no-fringe-field cases are non-physical
and such calculations should not be permitted in transport codes. The
aberrations in question are not caused by the fringe fields, but by the
broken symmetry inherent in a quadrupole of finite length.

We start with the quadrupole Hamiltonian and find canonical transformations
for both the electrostatic and magnetic cases which eliminate the
derivatives of the quadrupole strength up to 4$^{\rm th}$ order. In this
way, we easily reproduce the known aberration formulas, but with additional
physical insight.

\section{Theory}
\subsection{Electrostatic}
When using the longitudinal position as independent variable, the
Hamiltonian $H$ is just the longitudinal momentum:
\begin{equation}
  \label{unnH}
  H=-\sqrt{p_0^2-2mq\Phi-p_x^2-p_y^2}.
\end{equation}
The electrostatic potential is $\Phi(x,y,z)$ and $p_0$ is the reference
momentum. We will benefit from cleaner and more transparent notation if
momenta are measured in units of $p_0$. This has the additional benefit
that to first order, $p_x=x'$. Additionally, we let the potential $\Phi$ in
units of the reference kinetic energy $p_0^2/(2m)$. Then
\begin{equation}
  \label{H1}
  H=-\sqrt{1-\Phi-p_x^2-p_y^2}.
\end{equation}
We expand the square root to 4${\rm th}$ order in coordinates and ignore
the constant:
\begin{equation}
  \label{H2}
  H\approx {1\over 2}(\Phi+p_x^2+p_y^2)+{1\over 8}(\Phi+p_x^2+p_y^2)^2.
\end{equation}

To the same order, Laplace's equation gives for the expansion of the
quadrupole potential: 
\begin{equation}
  \label{V1}
  \Phi=V(z)(x^2-y^2)-{V''(z)\over 12}(x^4-y^4).
\end{equation}
The final Hamiltonian, correct to 4$^{\rm th}$ order is
\begin{eqnarray}
  \label{H3}
  H&=& {1\over 2}\left[V\,(x^2-y^2)-{V''\over
      12}(x^4-y^4)+p_x^2+p_y^2\right]\nonumber \\
&&+{1\over 8}\left[V\,(x^2-y^2)+p_x^2+p_y^2\right]^2.
\end{eqnarray}

The trouble with applying this to simple cases like thin lenses and
hard-edge limits is the presence of $V''(z)$, which becomes singular in
those limits. In most cases, one sacrifices physical insight and simply
traces particles with this Hamiltonian, using a more-or-less realistic
function $V(z)$. For example, the approach taken in {\tt GIOS}\cite{mw} is
to leave it up to the user to specify `fringe field integrals' such as
$\int V^2dz$ through the fringe fields. However, this leaves much room for
error; different integrals may not be realistic or consistent with each
other.  Moreover, if one needs to solve Laplace's equation to find fringe
field integrals, one might as well use the solution directly in a
ray-tracing code. If one does go through this exercise, one discovers that
the higher order aberrations are relatively insensitive to the `hardness'
of the quadrupole edges. This leads one to suspect that the aberrations are
dominated by an intrinsic effect which has nothing to do with the detailed
shape of the fringing field. Such is indeed the case.

It turns out to be possible to find a canonical transformation which
eliminates the derivatives of $V(z)$. In our case, we wish to
retain terms to 4$^{\rm th}$ order in the Hamiltonian (3$^{\rm rd}$ order
on force), and the transformation $(x,p_x,y,p_y)\rightarrow (X,P_X,Y,P_Y)$
has generating function
\begin{eqnarray}
  \label{gen}
  &&G(x,P_X,y,P_Y)=xP_X+yP_Y+{V'\over 24}(x^4-y^4)+\nonumber \\
                &&-{V\over 6}(x^3P_X-y^3P_Y).
\end{eqnarray}
To the same order, this yields the transformation
\begin{eqnarray}
  \label{trae}
  x&=&X+{V\over6}X^3\nonumber \\
  p_x&=&P_X-{V\over2}X^2P_X+{V'\over6}X^3.
\end{eqnarray}
The $y$-transformation is obtained by replacing $x,p_x,X,P_X$ with
$y,p_y,Y,P_Y$ and $V$ with $-V$. Note that outside the quadrupole, the
transformed coordinates are the same as the original ones.

This yields the transformed Hamiltonian $H^\ast$:
\begin{eqnarray}
  \label{Hstar}
  H^\ast&=&{V\over 2}(X^2-Y^2)+{1\over 2}(P_X^2+P_Y^2)+\nonumber\\
&+&{1\over 8}(P_X^2+P_Y^2)^2-{V\over 4}(X^2+Y^2)(P_X^2-P_Y^2)\nonumber\\
&+&{7V^2\over 24}(X^4+Y^4)-{V^2\over 4}X^2Y^2.
\end{eqnarray}
We can identify the terms: the first two are the usual linear ones; the
third term is not related to the electric field (it is small and due to the
fact that $x'\neq p_x$ or, equivalently, $\tan\theta\neq\sin\theta$); the
4$^{\rm th}$ term is also small and arises because a particle going through the
quadrupole at an angle is inside the quad for slightly longer than one
which remains on axis. See ref.\,\cite{DN} for more complete physical
derivation of the individual terms. 

The dominating higher order terms are the last two terms in
eqn.\,\ref{Hstar}. Since there are no derivatives of $V$, we can directly
write down the aberrations in the thin-lens limit:
\begin{eqnarray}
  \label{tlens}
  \Delta p_x&=&{-1\over f^2L}\left({7\over 6}x^3-{1\over 2}xy^2\right),
\end{eqnarray}
with a similar expression for $\Delta p_y$.  $L$ and $f$ are the
quadrupole's effective length and focal length. The fractional focal error
is found by dividing by the linear part $\Delta_0p_x=-x/f$:
\begin{equation}
  \label{frace}
  {\Delta f_x\over f}={1\over fL}\left({7\over 6}x^2-{1\over 2}y^2\right)
\end{equation}
for $x$, and similarly for $y$.

\subsection{Magnetic}
\label{sec:mag}
In magnetic fields, the canonical momentum $\vec{p}$ contains the vector
potential $\vec{A}$ so that the time-based Hamiltonian is
\begin{equation}
  \label{magHt}
   H_\tau={1\over 2m}\left|\vec{p}-q\vec{A}\right|^2
\end{equation}
As before, we use the invariant $p_0\equiv\sqrt{2mH_\tau}$ to normalize the
momenta, convert to $z$ as independent variable, and expand the square
root, keeping terms up to 4$^{\rm th}$ order:
\begin{eqnarray}
  \label{magHz}
  H&\approx& -A_z+{1\over 2}\left[(p_x-A_x)^2+(p_y-A_y)^2\right]+\nonumber\\
&+&{1\over 8}\left[(p_x-A_x)^2+(p_y-A_y)^2\right]^2.
\end{eqnarray}
To this order, the vector potential for quadrupole strength $k(z)$ is
\begin{eqnarray}
  \label{vecpot}
  &&A_x=-{k'\over 4}xy^2,\;\; A_y= {k'\over 4}x^2y\\
  &&A_z=-{k \over 2}(x^2-y^2)+{k''\over 48}(x^4-y^4),\nonumber
\end{eqnarray}
and the Hamiltonian can be written:
\begin{eqnarray}
  \label{magHz2}
    H&=& {1\over 2}\left[k\,(x^2-y^2)-{k''\over
      24}(x^4-y^4)+p_x^2+p_y^2\right]+\nonumber \\
&+&{k'xy\over 4}(yp_x-xp_y)+{1\over 8}(p_x^2+p_y^2)^2.
\end{eqnarray}
The generating function which will eliminate derivatives of $k$ is
\begin{eqnarray}
  \label{magGen}
  &&G(x,P_X,y,P_Y)=xP_X+yP_Y+{k'\over 48}(x^4-y^4)+\nonumber \\
                &&-{k\over 12}\left[(x^3+3xy^2)P_X-(3x^2y+y^3)P_Y\right],
\end{eqnarray}
which, to the same order yields transformation
\begin{eqnarray}
  \label{tram}
  x&=&X+{k\over12}(X^3+3XY^2)\\
p_x&=&P_X-{k\over4}\left[(X^2+Y^2)P_X-2XYP_Y\right]+{k'\over12}X^3,\nonumber
\end{eqnarray}
and similarly for $(y,p_y)$.  The transformed Hamiltonian is
\begin{eqnarray}
\label{magH}
  H^\ast&=&{k\over 2}(X^2-Y^2)+{1\over 2}(P_X^2+P_Y^2)+\nonumber\\
&+&{1\over 8}(P_X^2+P_Y^2)^2-{k\over 4}(X^2+Y^2)(P_X^2-P_Y^2)\nonumber\\
&+&{k^2\over 12}(X^4+Y^4)+{k^2\over 2}X^2Y^2.
\end{eqnarray}
Notice the similarity to eqn.\,\ref{Hstar}: in fact all terms are identical
except the last two, which only differ in their coefficients. Applying the same
procedure as in the electrostatic case, we write down the fractional change
in focusing strength:
\begin{equation}
  \label{fracm}
  {\Delta f_x\over f}={1\over fL}\left({1\over 3}x^2+y^2\right)
\end{equation}

\section{Discussion}
\label{sec:dis}
Formulas \ref{frace} and \ref{fracm} are handy for quickly evaluating the
importance of 3$^{\rm rd}$ order aberration. They also show that for fixed
focal length, the {\bf only} way of reducing the aberration is by
lengthening the quadrupole; the fraction of aperture used is not important;
for a given effective length, the absolute size of the aperture is not
important; the shape of the ends of the electrode is not important.

Comparing the two formulas, we see that for roundish beams ($x\approx y$),
electrostatic and magnetic quads yield similar aberrations: they are in the
ratio of ${7\over6}:{4\over3}$. For cases where one transverse dimension is
large compared with the other, and it is important to maintain the quality
in the larger dimension, magnetic quads are better by a factor of
${7\over2}$.  However, for the more common case where it is more important
to maintain the quality of the higher quality dimension, electrostatic
quads win by a factor of 2.

Results from using the above Hamiltonians are in agreement with those from
using the commonly used codes {\tt GIOS} and {\tt COSY}, provided fringe
field cards are used. In both of those codes it is possible to perform a
3$^{\rm rd}$ order calculation with quads which have no fringe fields. This
gives incorrect and actually completely unphysical results. In essence,
omitting the fringe field cards in those codes describes a situation where
the particle traverses non-Maxwellian fields. For example, {\tt GIOS},
since it does not use the scalar value of the potential field, does not
obey conservation of energy when fringe field cards are omitted.

The hard-edge case is correctly described in {\tt GIOS} by including fringe
field cards and setting the quadrupole aperture to zero, or, equivalently,
setting all the fringe field integrals to zero. This is a useful
approximation since the results are usefully close to reality and yet one
needs not worry about specifying realistic fringe field integrals. This
does not work in {\tt COSY}, since a zero aperture forces an infinitesimal
integration step-size. A better solution would be to build in the hard-edge
kicks and use these as default when no fringe field is specified.

The required hard-edge kicks at the entrance to the quadrupole are derived
directly from equations \ref{trae} and \ref{tram}. The reason is that we
know that the transformed coordinates $(X,P_X,Y,P_Y)$ do not experience any
singular forces in the hard-edge limit. Therefore, the kicks for those
coordinates are all zero. So the kicks for the untransformed $(x,p_x,y,p_y)$
for the electrostatic case are,
\begin{eqnarray}
  \label{kicke}
  \Delta x&=& {V\over6}x^3\nonumber\\
\Delta p_x&=&{-V\over2}x^2p_x\nonumber\\
  \Delta y&=&{-V\over6}y^3\\
\Delta p_y&=& {V\over2}y^2p_y,\nonumber
\end{eqnarray}
and for the magnetic case are,
\begin{eqnarray}
  \label{kickm}
  \Delta x&=& {k\over12}(x^3+3xy^2)\nonumber\\
\Delta p_x&=&{-k\over 4}\left[(x^2+y^2)p_x-2xyp_y\right]\nonumber\\
  \Delta y&=&{-k\over12}(3x^2y+y^3)\\
\Delta p_y&=&{ k\over 4}\left[(x^2+y^2)p_y-2xyp_x\right].\nonumber
\end{eqnarray}
The kicks at the exit are, of course, opposite in sign. These agree with
the {\tt GIOS} case of zero fringe field integrals. See ref.\,\cite{mw}.

\end{document}